\documentstyle[preprint,eqsecnum,aps]{revtex}
\def\eps{\varepsilon}
\def\D{\widetilde{\cal D}}
\def\const{{\rm const\,}}
\def\partt{\mbox{\boldmath $\partial$}}

\begin{document}
\draft
\preprint{SPbU IP-98-4, 13 January 98, chao-dyn/9801033, 27 January 98}
\title{Renormalization Group, Operator Product Expansion,
and Anomalous Scaling in a Model of Advected Passive Scalar}
\author{ Loran Ts. Adzhemyan}
\author{ Nikolaj V. Antonov}
\author{ Alexander N. Vasil'ev}
\address{ Department of Theoretical Physics, St Petersburg University, \\
Uljanovskaja 1, St Petersburg---Petrodvorez, 198904, Russia}
%\date{}
\maketitle
\begin{abstract}
Field theoretical renormalization group methods
are applied to the Obukhov--Kraichnan model of a passive scalar
advected by the Gaussian velocity field with the covariance
\[\langle{\bf v}(t,{\bf x}){\bf v}(t',{\bf x})\rangle -
\langle{\bf v}(t,{\bf x}){\bf v}(t',{\bf x'})\rangle
\propto\delta(t-t')|{\bf x}-{\bf x'}|^{\eps}.\]
Inertial range anomalous scaling for the structure functions and
various pair correlators is established as a consequence of
the existence in the corresponding operator product expansions
of ``dangerous'' composite opera\-tors [powers of the local
dissipation rate], whose {\it negative} critical dimensions
determine anomalous exponents.
The main technical result is the calculation of the anomalous
exponents in the order $\eps^{2}$ of the $\eps$ expansion.
Generalization of the results obtained to the case of a ``slow''
velocity field is also presented.
\end{abstract}

\baselineskip17.5pt
\section{Introduction}
\label {sec-0}

One of the main problems in the modern theory of fully
developed turbulence is to verify the basic principles of the
Kolmogorov--Obukhov (KO) phenomenological theory
\cite{Monin,Orszag} within the framework of a microscopic model
and to investigate deviations from this theory, provided they
exist.

In particular, one is interested in the single-time ``structure
functions''
\begin{equation}
S_{n}(r)\equiv\langle[\theta({\bf x})-\theta({\bf x'})]^{n}\rangle,
\quad r\equiv|{\bf x}-{\bf x'}|
\label{A1}
\end{equation}
in the inertial range.
Here $\theta(x)\equiv\theta(t,{\bf x})$ can be the component
of the velocity field directed along the vector ${\bf x}-{\bf x'}$,
or the scalar field in the problem of turbulent advection;
the brackets  $\langle \dots\rangle$ denote the ensemble averaging,
and the time argument common to all the quantities is omitted
in (\ref{A1}) and analogous formulas below.
The inertial range (or the convective range in the problem of
turbulent advection)  is defined by the inequalities
$l<<r<<L$, where $l\equiv\Lambda^{-1}$ is the internal (viscous)
scale and $L\equiv M^{-1}$ is the external (integral) scale.

According to the KO theory, the functions  (\ref{A1}) in the
inertial range are independent of both the external and internal
scales (the First and the Second Kolmogorov hypotheses, respectively)
and are determined by the only parameter $\bar{\epsilon}$, the mean
dissipation rate, see e.g. \cite{Monin,Orszag}.
Dimensionality considerations then determine the functions
(\ref{A1}), apart from numerical coefficients, in the form
\begin{equation}
S_{n}(r)=\const (\bar{\epsilon}r)^{n/3}.
\label{A2}
\end{equation}

Both experimental and theoretical evidence is known in favor
of some deviation from the predictions of the KO theory, see
\cite{Monin,33,34,Combustion,Menev,86}.
For the structure functions
(\ref{A1}), these deviations are phenomenologically written
in the form
(in contradiction with the First Kolmogorov hypothesis)
\begin{equation}
S_{n}(r)=\const (\bar{\epsilon}r)^{n/3}\, f_{n}(Mr),
\label{A3}
\end{equation}
where the ``scaling functions'' $f_{n}(Mr)$
are supposed to have power-like behaviour in the
asymptotic region $Mr<<1$,
\begin{equation}
 f_{n}(Mr)\simeq\const (Mr)^{q_{n}}.
\label{A4}
\end{equation}

The singular dependence of the structure functions on $M$
for $M\to0$ and nonlinearity of the exponents $q_{n}$ in $n$ are
usually referred to as ``anomalous scaling'', and in
theoretical models they are explained by strongly developed
fluctuations of the local dissipation rate (``intermittency''),
see discussion in \cite{Monin}, Chap.25. Within the framework of
numerous models, the anomalous exponents $q_{n}$  are related
to the statistics of the local dissipation rate or to
the dimensionality of fractal structures formed by small-scale
vortices in the dissipative range, see for a review \cite{86}.
As a rule, these models are only loosely related to underlying
microscopic models, and therefore some doubt remains about
the universality of representations like (\ref{A3}), (\ref{A4})
and the very existence of deviations from the KO theory.

An effective method of studying self-similar scaling behaviour
is that of the renormalization group (RG), see \cite{Bogol,Collins}.
It was successfully applied in the theory of
critical phenomena to explain the origin of critical scaling and
to calculate universal quantities (critical dimensions and
scaling functions) in the form of  series in the formal small
parameter $\eps=4-d$, where $d$ is the space dimensionality,
see \cite{Zinn}. This technique is also fully applicable to
the theory of turbulence \cite{DeDom2}, see also the review
paper \cite{UFN} and references therein.

In the  statistical theory of turbulence,  the microscopic
model is usually taken to be the stochastic Navier--Stokes
equation (SNS) with an external random force which imitates
the injection of energy by large-scale modes,
see e.g. \cite{Monin,Orszag,UFN}.
The role of the RG expansion parameter
is played by the exponent $\eps$ entering into the random force
correlator.
For the structure functions (\ref{A1}), the RG method allows
proof of the Second Kolmogorov hypothesis (independence of
the viscous length) for a wide variety of realistic random
forces, see \cite{UFN,JETP,AV}. This
is equivalent to the representation (\ref{A3}), in which the
form of the scaling function $f_{n}(Mr)$ is arbitrary. We note
that the representations (\ref{A3}) for any functions $f_{n}(Mr)$
imply the existence of  scaling in the infrared (IR) region
$L,r>>l$ with definite ``critical dimensions''
\begin{equation}
\Delta_{M}=-\Delta_{r}=1,\quad \Delta[S_{n}]=-n/3
\label{A5}
\end{equation}
for any fixed value of $\bar{\epsilon}$.
This means that the structure
functions (\ref{A3}) scale as $S_{n}\to \lambda^{\Delta[S_{n}]}
S_{n}$ upon the substitution $M\to \lambda^{\Delta_{M}} M$,
$r\to \lambda^{\Delta_{r}} r$ [in general, the exponent
$\Delta[S_{n}]$ is replaced by the critical dimension of the
corresponding correlation function; this dimension is calculated
within the  $\eps$ expansion]. It is this IR scaling that is
analogous to the critical scaling in the theory of critical
phenomena, where the role of an external scale is played by the
correlation length $L\equiv r_{c}$, see  \cite{Zinn}.
In the RG approach, the critical dimensions like (\ref{A5})
arise as coefficients in the RG equations. The form of the
scaling functions is not determined by the RG equations themselves,
and therefore the anomalous exponents in (\ref{A4}) are {\it not}
related to the critical dimensions of the functions (\ref{A1}).

As in the case of critical phenomena \cite{Zinn}, the dependence of
the scaling functions on the argument $Mr$ in the region  $Mr<<1$
is studied here using the well known operator product expansion
(OPE, or SDE=short distance expansion), see [12--17].
According to the OPE, the scaling
functions in expressions like (\ref{A3}) are represented in the form
\begin{equation}
f(Mr) = \sum_{F} C_{F} (Mr)^{\Delta_{F}},\quad Mr\to0.
\label{A6}
\end{equation}
Here $C_{F}$ are coefficients regular in $M$, the summation
is implied over all possible composite operators $F$
allowed by the symmetry of the left-hand side, and
$\Delta_{F}$ are their critical dimensions.
In particular, only scalar Galilean invariant operators with nonzero
mean value contribute to the OPE for the structure functions
(\ref{A1}), see  \cite{UFN,Kim}.

In the theory of critical phenomena, all the nontrivial composite
operators have positive critical dimensions, $\Delta_{F}>0$, and
the leading term in (\ref{A6}) is determined by the simplest
operator $F=1$ with $\Delta_{F}=0$, i.e., the function $f(Mr)$
is finite as $M\equiv r_{c}^{-1}\to0$, see \cite{Zinn}.
It has long been realized \cite{JETP} (see also papers
[12,14--17]) that the singular behaviour of the scaling
functions in the SNS model for $M\to0$  is related to the
existence in the model of composite operators with
{\it negative} critical dimensions. These operators, called
``dangerous'' in \cite{JETP}, should not be confused with IR
relevant operators considered in \cite{Eyink2,Relevant},
whose existence can lead to the violation of the scaling regime.

Dangerous composite operators in the SNS model occur only for
finite values of the RG expansion parameter $\eps$, and within
the $\eps$ expansion it is impossible to decide whether or not
a given operator is dangerous, provided its critical dimension
is not found exactly using the Schwinger-type functional equations
or the Galilean symmetry, see [12,15--17,19--21].
Moreover, dangerous operators enter into the operator product
expansions in the form of infinite families with the
spectrum of critical dimensions unbounded from below, and
the analysis of the small $M$ behaviour implies the summation
of their contributions. Such a summation for the
case of different-time correlators, first accomplished
in \cite{JETP} (see also \cite{UFN,Kim}), establishes
the substantial $M$  dependence of the correlators and their
superexponential decay as the time differences increase.
Of course, this effect is well-known and has a simple physical
interpretation as the transport of turbulent eddies as a whole
by the large-scale modes, see \cite{Kraich4}. Therefore, this
effect is not ignored within the correct RG formalism, as it is
sometimes believed \cite{Chen,Woodruff}, provided the
RG is used beyond the $\eps$ expansion and is combined with
the OPE technique.

The analysis of the  $M$  dependence of Galilean invariant
quantities, such as the single-time structure functions
(\ref{A1}), requires the explicit construction of all dangerous
invariant scalar operators, exact calculation of their critical
dimensions, and summation of their contributions in the corresponding
operator product expansions (\ref{A6}).  This is clearly not a simple
problem and it requires considerable improvement of the
present technique.

In view of the difficulties encountered by the RG approach to
the SNS model it is tempting to apply the formalism to simpler
models, which exhibit some of the features of real turbulent
flows, but are easier to study. An interesting example is
provided by the well-known Heisenberg model, see \cite{Monin},
Chap.17, whose exact solution has recently been rederived
using the RG method \cite{Heis}. Unfortunately, the  Heisenberg
model does not involve higher-order functions and is therefore
not suitable for studying anomalous scaling.

Recently, much attention has been attracted by a simple model
of the passive advection of a scalar quantity by a Gaussian
velocity field, introduced by Obukhov \cite{Ob} and Kraichnan
\cite{Kraich1}, see the papers [28--40] and
references therein. It turns out, that the structure functions of
the scalar field in this model exhibit anomalous scaling behaviour
analogous to (\ref{A3}), (\ref{A4}), and the corresponding anomalous
exponents can be calculated explicitly within  expansions in
certain small parameters, [31--34].

The advection of a passive scalar field $\theta(x)\equiv
\theta(t,{\bf x})$ is described by the stochastic equation
\begin{equation}
\nabla _t\theta=\nu _0\triangle \theta+f , \quad
\nabla _t\equiv \partial _t+ v_{i}\, \partial_{i},
\label{1}
\end{equation}
where $\partial _t \equiv \partial /\partial t$,
$\partial _i \equiv \partial /\partial x_{i}$, $\nu _0$
is the molecular diffusivity coefficient, $\triangle$
is the Laplace operator, ${\bf v}(x)$ is the transverse (owing
to the incompressibility) velocity field, and $f\equiv f(x)$ is a
Gaussian scalar noise with zero mean and correlator
\begin{equation}
\langle  f(x)  f(x')\rangle = \delta(t-t')\, C(Mr), \quad
r\equiv|{\bf x}-{\bf x'}|.
\label{2}
\end{equation}
The parameter $L\equiv M^{-1}$ is an integral scale related
to the scalar
noise, and $C(Mr)$ is some function finite as $L\to\infty$.
Without loss of generality, we take $C(0)=1$ (the
dimensional coefficient in  (\ref{2}) can be absorbed by
appropriate rescaling of the field $\theta$ and noise $f$).

In a more  realistic formulation, the field  ${\bf v}(x)$
satisfies the stochastic Navier--Stokes equation, see e.g.
\cite{Passive}.  Following [26--34],
we shall consider a simplified model in
which  ${\bf v}(x)$ obeys a Gaussian distribution with zero
average and correlator
\begin{equation}
\langle v_{i}(x) v_{j}(x')\rangle = D_{0}\,
\frac{\delta(t-t')}{(2\pi)^d}
\int d{\bf k}\, P_{ij}({\bf k})\, (k^{2}+m^{2})^{-d/2-\eps/2}\,
\exp [{\rm i}{\bf k}\cdot({\bf x}-{\bf x'})] ,
\label{3}
\end{equation}
where $P_{ij}({\bf k}) = \delta _{ij} - k_i k_j / k^2$ is the
transverse projector, $k\equiv |{\bf k}|$, $D_{0}$ is an amplitude
factor, $1/m$ is another integral scale, and $d$  is the
dimensionality of the ${\bf x}$  space; $0<\eps<2$ is a parameter
with the real (``Kolmogorov'') value $\eps=4/3$;
$\eps=2-\gamma$ in the notation of \cite{Falk1,Falk2},
$\eps=\kappa$ in the notation of \cite{GK}, and $\eps=\chi$
in the notation of \cite{BGK}.
The relations
\begin{equation}
g_{0} \equiv D_{0}/\nu_0 \equiv \Lambda^{\eps}
\label{g0}
\end{equation}
define the coupling constant $g_{0}$ (expansion parameter in the
perturbation theory, see Sec.\ref{sec-1}) and the characteristic
UV momentum scale $\Lambda$ ($\Lambda\simeq 1/r_{d}$ in the
notation of \cite{Falk1,Falk2}).

In the model (\ref{1})--(\ref{3}), the odd multipoint correlation
functions of the scalar field vanish, while the even single-time
functions satisfy linear partial differential equations.
The solution for the pair correlator is obtained
explicitly; it shows that the structure function $S_{2}$ is
finite for $M,m=0$ \cite{Kraich1}. The higher-order
correlators are not found explicitly, but their asymptotic
behaviour for $M\to0$ can be extracted from the analysis of
the nontrivial zero modes of the corresponding differential
operators in the limit $1/d\to0$ \cite{Falk1,Falk2}
or $\eps\to0$ \cite{GK,BGK}. It was shown that
the structure functions are finite for $m=0$, and in the
convective range $\Lambda >>1/r>>M$ they have the form
(up to the notation):
\begin{equation}
S_{2n}(r)\equiv\langle[\theta({\bf x})-\theta({\bf x'})]^{2n}
\rangle\propto D_{0}^{-n}\, r^{n(2-\eps)}\, (Mr)^{\Delta_{n}},
\label{HZ1}
\end{equation}
with negative anomalous exponents $\Delta_{n}$, whose first
terms of the expansion in $1/d$ \cite{Falk1,Falk2} and
$\eps$ \cite{GK,BGK} have the form
\begin{equation}
\Delta_{n}=-2n(n-1)\eps/(d+2)+O(\eps^{2})=-2n(n-1)\eps/d +O(1/d^{2}).
\label{HZ3}
\end{equation}
We note that $\Delta_{n}=-\rho_{2n}$ in the notation of \cite{GK},
$\Delta_{2}=-\Delta$ in the notation of \cite{Falk1}, and
$\Delta_{n}=-\Delta_{2n}$ in \cite{Falk2}.

Another quantity of interest is the local dissipation rate,
$E(x)=\nu_0 [\partial_{i}\theta(x)\partial_{i}\theta(x)]$.
The single-time correlation functions of its powers in the
convective range have the form
\cite{Falk2,BGK}:
\begin{equation}
\langle E^{n}(x)\,E^{m}(x')\rangle
\propto (\Lambda r)^{-\Delta_{n}-\Delta_{m}} (Mr)^{\Delta_{n+m}}
\label{HZ2}
\end{equation}
where $\Lambda$ is defined in (\ref{g0})
and the exponents $\Delta_{n}$ are the same as in Eq.(\ref{HZ1}).

In this paper, we apply the RG and OPE approach developed in
[12--17,19-21,41]
to the model (\ref{1})--(\ref{3}). We show that the RG explains
the origin of anomalous scaling and allows the anomalous
exponents to be calculated in the form of a series in the
parameter $\eps$  entering into the correlator (\ref{2}). Therefore,
this parameter plays in the RG approach the role analogous to
that played by the parameter $\eps=4-d$ in the RG
theory of critical phenomena, and the results obtained in
\cite{GK,BGK} can be interpreted as the first terms of
the corresponding $\eps$ expansions.

We have calculated the
exponents $ \Delta_{n}$  in the second order of the
 $\eps$ expansion for an arbitrary value of the space
dimensionality $d$ (they are given in Sec.\ref{sec-2}).
In particular, we have obtained
\begin{eqnarray}
\Delta_{n}=-\frac{n(n-1)\eps}{2}
+\frac{n(n-1)\eps^{2}}{16}
\left[ 2n(19-66\ln(4/3))+5-24 \ln(4/3) \right]
+O(\eps^{3})= \nonumber \\
= - n(n-1)\eps/2 + \eps^{2} n(n-1)
[0.00162n-0.11902] +O(\eps^{3})
\label{d=2}
\end{eqnarray}
for $d=2$  and
\begin{eqnarray}
\Delta_{n}=-\frac{2n(n-1)\eps}{5}+
\frac{2n(n-1)\eps^{2}}{875} \times
\nonumber \\
\times\left[ 2n(255\pi\sqrt{3}-1384)+345\pi\sqrt{3}
-1884\right]+O(\eps^{3})=
\nonumber \\
= - 2n(n-1)\eps/5 + \eps^{2} n(n-1)
[0.01626n-0.01535] +O(\eps^{3})
\label{d=3}
\end{eqnarray}
for $d=3$.

Representations like (\ref{HZ1}), (\ref{HZ2}) can also be
derived using the RG and OPE for any correlation function;
as an additional example, we consider the single-time pair
correlators involving  second-rank irreducible tensors
of the form
\[\partial_{i}\theta\partial_{j}\theta
[\partial_{s}\theta\partial_{s}\theta]^{n-2}-\delta_{ij}
[\partial_{s}\theta\partial_{s}\theta]^{n}/d,\]
the special
case $n=2$ was considered previously in \cite{Falk2}.

From the RG viewpoints, the model (\ref{1})--(\ref{3}) is
simpler than the ``true'' SNS model in at least two respects.
First, dangerous operators in the model  (\ref{1})--(\ref{3})
exist even for asymptotically small values of  $\eps$ and
therefore they can be identified within the $\eps$ expansion.
These are the local dissipation rate $E(x)$  and all its powers,
their critical dimensions being $\Delta_{n}$.  Second,
only finite number of these operators contribute to the operator
product expansion (\ref{A6}) for any given correlation function,
and the additional resummation of the series (\ref{A6}), discussed
above, is not required here:  the leading term of the asymptotic
behaviour of the function $f(Mr)$ for $Mr\to0$ is simply given by
the contribution of the ``most dangerous'' operator, i.e., that
having the smallest value of $\Delta_{n}$.

It was suggested in \cite{GK} that the anomalous exponents
describing the asymptotic behaviour for $Mr<<1$  can be related
to certain IR relevant interactions in a hypothetical
RG approach, see  also discussion in \cite{Falk2}.
This is not exactly so: in fact,
these exponents in the model (\ref{1})--(\ref{3}) are
related to the critical dimensions of certain ``dangerous''
composite operators in the operator product expansions for
corresponding Green functions. It should be stressed, that in
the SNS model the behaviour at  $Mr\to0$ will probably be
determined by additional resummations of the operator
product expansions, which can lead to more exotic behaviour
than the simple power-like one in  (\ref{HZ1}), (\ref{HZ2}).

The plan of our paper is the following.
In Sec.\ref{sec-1}, we discuss the field theoretical formulation
and UV renormalization of the model (\ref{1})--(\ref{3}) and
derive the corresponding RG equations with exactly known RG
functions (the $\beta$ function and the anomalous dimension).
These equations have an IR stable fixed point, which establishes
the existence of IR scaling with exactly known critical
dimensions of the basic fields and parameters of the model.
In Sec.\ref{sec-2}, we discuss the renormalization of various
composite operators in the model (\ref{1})--(\ref{3}). In
particular, we present the second-order result for the $\eps$
expansion of the quantity $\Delta_{n}$, the critical dimension
associated with the operator $E^{n}(x)$. In Sec.\ref{sec-4},
it is explained using the OPE that, at the same time, $\Delta_{n}$
play the part of the anomalous exponents in relations like
(\ref{HZ1}), (\ref{HZ2}). The results obtained are discussed
in Sec.\ref{sec-5}. We also briefly mention there one of the
possible modifications of the model  (\ref{1})--(\ref{3}),
that of a ``slow'' velocity field, in which the correlator
(\ref{3}) contains no $\delta$ function in time.
The RG analysis shows that the Green
functions in this model exhibit an anomalous scaling behaviour,
and the corresponding anomalous exponents are calculated
in the form of series in $\eps+2$.

\section{Field theoretical formulation, UV
re\-nor\-ma\-li\-za\-ti\-on, and RG equations}
\label {sec-1}

The stochastic problem (\ref{1})--(\ref{3}) is equivalent
to the field theoretical model of the set of three fields
$\Phi\equiv\{\theta, \theta',{\bf v}\}$ with action functional
\begin{equation}
S(\Phi)=\theta' D_{\theta}\theta' /2+
\theta' \left[ - \partial_{t}\theta -({\bf v}\partt) \theta
+ \nu _0\triangle \theta \right]
-{\bf v} D_{v}^{-1} {\bf v}/2.
\label{10}
\end{equation}
The first four terms in  (\ref{10}) represent
the Martin--Siggia--Rose-type action  [42--45]
for the stochastic problem (\ref{1}), (\ref{2}) at fixed ${\bf v}$,
and the last term
represents the Gaussian averaging over ${\bf v}$. Here $D_{\theta}$
and $D_{v}$ are the correlators (\ref{2}) and (\ref{3}), respectively,
the required integrations over $x=(t,{\bf x})$ and summations over
the vector indices are understood.

The formulation (\ref{10}) means that statistical averages
of random quantities in stochastic problem  (\ref{1})--(\ref{3})
coincide with functional averages with the weight $\exp S(\Phi)$,
therefore generating functionals of total ($G(A)$) and connected
($W(A)$) Green functions of the problem
are represented by the functional integral
\begin{equation}
G(A)=\exp  W(A)=\int {\cal D}\Phi \exp [S(\Phi )+A\Phi ]
\label{14}
\end{equation}
with arbitrary sources $A\equiv A^{\theta},A^{\theta'},A^{\bf v}$
in the linear form
\[A\Phi \equiv \int dx[A^{\theta}(x)\theta (x)+A^{\theta '}(x)\theta '(x)
+ A^{\bf v}_{i}(x)v_{i}(x)].\]

The model (\ref{10}) corresponds to a standard Feynman
diagrammatic techni\-que with the triple vertex
$-\theta'({\bf v}\partt)\theta$
and bare propagators (in the momentum-frequency representation)
\begin{eqnarray}
\langle \theta \theta' \rangle _0=\langle \theta' \theta \rangle _0^*=
(-i\omega +\nu _0k^2)^{-1} ,   \nonumber
\\
\langle \theta \theta \rangle _0=C(k)\,(\omega ^2+\nu _0^2k^4)^{-1},
\nonumber \\
\langle \theta '\theta '\rangle _0=0 ,
\label{16}
\end{eqnarray}
where $C(k)$ is the Fourier transform of the function $C(Mr)$ in
(\ref{2}) and the bare propagator $\langle{\bf v}{\bf v}\rangle _0$
is given by Eq.(\ref{3}). The role of the coupling constant in the
perturbation theory is played by the parameter
$g_{0}$ defined in (\ref{g0}).

It is well known that the analysis of UV divergences is based on
the analysis of canonical dimensions. Dynamical models of the type
(\ref{10}), in contrast to static models, are two-scale,
i.e., to each quantity $F$ (a field or a parameter in the action
functional) one can assign two independent canonical dimensions,
the momentum dimension $d_{F}^{k}$ and the frequency dimension
$d_{F}^{\omega}$, determined from the natural normalization
conditions
\[d_k^k=-d_{\bf x}^k=1,\ d_k^{\omega }=d_{\bf x}^{\omega }=0 ,
d_{\omega }^k=d_t^k=0,\ d_{\omega }^{\omega }=-d_t^{\omega }=1, \]
and from the requirement that each term of the action functional
be dimensionless (with respect to the momentum and frequency
dimensions separately), see \cite{UFN,Pismak}.
Then, based on $d_{F}^{k}$ and $d_{F}^{\omega}$, one can
introduce the ``summed'' (total) canonical dimension
$d_{F}=d_{F}^{k}+2d_{F}^{\omega}$ (in the free theory,
$\partial_{t}\propto\triangle$).

The dimensions for the model (\ref{10}) are given in
Table \ref{table1}, including renormalized parameters,
which will be considered later on.

From Table \ref{table1} it follows that the model is
logarithmic (the
coupling constant $g_{0}$  is dimensionless) at $\eps=0$,
and the UV divergences have the form of the poles in $\eps$
in the Green functions.

The total dimension $d_{F}$ plays in the theory of
renormalization of dynamical models the same role as does
the conventional (momentum) dimension in static problems.
The canonical dimensions of an arbitrary
1-irreducible Green function $\Gamma = \langle\Phi \dots \Phi
\rangle _{\rm 1-ir}$ are given by the relations
\begin{eqnarray}
d_{\Gamma }^k=d- N_{\Phi}d_{\Phi},\nonumber \\
d_{\Gamma }^{\omega }=1-N_{\Phi }d_{\Phi }^{\omega },\nonumber \\
d_{\Gamma }=d_{\Gamma }^k+2d_{\Gamma }^{\omega }=
d+2-N_{\Phi }d_{\Phi},
\label{17}
\end{eqnarray}
where $N_{\Phi}=\{N_{\theta},N_{\theta'},N_{\bf v}\}$ are the
numbers of corresponding fields entering into the function
$\Gamma$, and the summation over all types of the fields is
implied.
The total dimension $d_{\Gamma}$ is the formal index of the
UV divergence. Superficial UV divergences, whose removal requires
counterterms, can be present only in those functions $\Gamma$ for
which $d_{\Gamma}$ is a nonnegative integer, see e.g.
\cite{Bogol,Collins}.

Analysis of divergences should be based on the following auxiliary
considerations:

(1) From the explicit form of the vertex and bare propagators in
the model (\ref{10}) it follows that $N_{\theta'}- N_{\theta}=2N_{0}$
for any 1-irreducible Green function, where $N_{0}\ge0$
is the total number of the bare propagators $\langle \theta \theta
\rangle _0$ entering into the function (obviously, no diagrams
with $N_{0}<0$ can be constructed). Therefore, the difference
$N_{\theta'}- N_{\theta}$ is an even nonnegative integer for
any nonvanishing function.

(2) If for some reason a  number of external momenta occur as an
overall factor in all the diagrams of a given Green function, the
real index of divergence $d_{\Gamma}'$ is smaller than $d_{\Gamma}$
by the
corresponding number of unities (the Green function requires
counterterms only if $d_{\Gamma}'$  is a nonnegative integer).

In the model (\ref{10}), the derivative $\partial$ at the
vertex $\theta'({\bf v}\partt)\theta$ can be moved onto the
field $\theta'$ by virtue of the transversality of the field
${\bf v}$. Therefore, in any 1-irreducible diagram it is always
possible to move the derivative onto any of the external
``tails'' $\theta$ or $\theta'$, which decreases the real index
of divergence: $d_{\Gamma}' = d_{\Gamma}- N_{\theta}-N_{\theta'}$.
The fields $\theta$, $\theta'$ enter into the counterterms only
in the form of derivatives $\partial\theta$, $\partial\theta'$.

From the dimensions in Table \ref{table1} we find
$d_{\Gamma} = d+2 - N_{\bf v}
+ N_{\theta}- (d+1)N_{\theta'}$  and $d_{\Gamma}'=(d+2)(1-N_{\theta'})
- N_{\bf v}$. From these expressions it follows that for any $ d$,
superficial divergences can exist only in the 1-irreducible functions
$\langle\theta'\theta\dots\theta\rangle$ with $N_{\theta'}=1$
and arbitrary value of $N_{\theta}$, for which $d_{\Gamma}=2$,
$d_{\Gamma}'=0$. However, all the functions with $N_{\theta}>
N_{\theta'}$ vanish (see above) and obviously do not require
counterterms. We are left with the only superficially
divergent function $\langle\theta'\theta\rangle$;
the corresponding counterterm must contain two symbols
$\partial$  and is therefore reduced to $\theta'\triangle\theta$.
Inclusion of this counterterm is reproduced by the multiplicative
renormalization of the parameters $g_{0},\nu_0$ in the action
functional (\ref{10}) with the only
independent renormalization constant $Z_{\nu}$:
\begin{equation}
\nu_0=\nu Z_{\nu},\quad g_{0}=g\mu^{\eps}Z_{g},
\quad Z_{g}=Z_{\nu}^{-1}.
\label{18}
\end{equation}
Here $\mu$ is the renormalization mass in the minimal subtraction
scheme (MS), which we always use in what follows, $g$ and $\nu$
are renormalized analogues of the bare parameters $g_{0}$ and $\nu_0$,
and $Z=Z(g,\eps,d)$ are the renormalization constants. Their
relation in (\ref{18}) results from the absence of renormalization
of the contribution with $D_{0}$ in (\ref{10}), so that
$D_{0}\equiv g_{0}\nu_0 = g\mu^{\eps} \nu$, see (\ref{g0}).
No renormalization of the fields and ``masses'' is required,
i.e., $Z_{\Phi}=1$ for all $\Phi$ and  $m_{0}=m$, $M_{0}=M$,
$Z_{m}=Z_{M}=1$.

Since the fields are not renormalized, their renormalized Green
functions $W^{R}$ coincide with the corresponding unrenormalized
functions $W=\langle\Phi\dots\Phi\rangle$ (for definiteness, we
discuss the connected functions); the only difference is
in the choice of variables and in the form of perturbation theory
(in $g$ instead of $g_{0}$):
\begin{equation}
W^{R} (g,\nu,\mu,\dots) = W (g_{0},\nu_0,\dots)
\label{20}
\end{equation}
(the dots stand for other arguments like coordinates and momenta).
We use $\D_{\mu}$ to denote the differential operator
$\mu\partial_{\mu}$ for fixed bare parameters $g_{0},\nu_0$
and operate on both sides of Eq.(\ref{20})  with it. This
gives the basic differential RG equation:
\begin{eqnarray}
{\cal D}_{RG}W^{R} (g,\nu,\mu,\dots) = 0, \nonumber \\
{\cal D}_{RG}\equiv {\cal D}_{\mu} + \beta(g)\partial_{g}
-\gamma_{\nu}(g){\cal D}_{\nu},
\label{200}
\end{eqnarray}
where we have written ${\cal D}_{s}\equiv s\partial_{s}$ for
any variable $s$, and the RG functions (the $\beta$ function and
the anomalous dimension $\gamma$) are defined as
\begin{equation}
\gamma_{\nu}(g)\equiv\D_\mu \ln Z_{\nu},\quad
\beta(g)\equiv \D_\mu g= g[-\eps + \gamma_{\nu}].
\label{21}
\end{equation}
The relation between $\beta$ and $\gamma$ results from the
definitions and the last relation in (\ref{18}). In general, if
some quantity $G$ is renormalized multiplicatively,
$G=Z_{G}G^{R}$, it satisfies the RG equation of the form
\begin{equation}
\left[{\cal D}_{RG}+ \gamma_{G}(g)\right]G^{R}=0, \quad
\gamma_{G}(g)\equiv \D_\mu \ln Z_{G}
\label{22}
\end{equation}
with the operator ${\cal D}_{RG}$ from (\ref{200}).

Explicit calculation of the constant $Z_{\nu}$ in the model
(\ref{10}) in the one-loop approximation gives:
\begin{equation}
Z_{\nu}= 1-\frac{g\,(d-1)\, C_{d}}{2d\eps},
\label{25}
\end{equation}
where we have written
$C_{d} \equiv  {S_{d}}/{(2\pi)^{d}}$
and $S_d\equiv 2\pi ^{d/2}/\Gamma (d/2)$
is the surface area of the unit sphere in $d$-dimensional space.

The one-loop approximation (\ref{25}) for the constant $Z_{\nu}$
is in fact exact, i.e., it has no corrections of order $g^{2}$,
$g^{3}$, and so on. Indeed, from the explicit form of the vertex and
the bare propagators (\ref{3}), (\ref{16}) it follows that any
multiloop diagram of the 1-irreducible function $\langle\theta'
\theta\rangle$ contains effectively a closed circuit of retarded
propagators $\langle\theta\theta'\rangle_{0}$  and therefore
vanishes (it is also important here that the propagator $\langle
{\bf v}{\bf v}\rangle_{0}$ in (\ref{3}) is proportional to the
$\delta$ function in time).

From the definitions (\ref{21}) using Eq.(\ref{25}) we find exact
expressions for the basic RG functions:
\begin{equation}
\gamma_{\nu}(g)=\frac{g(d-1)C_{d}}{2d},\quad
\beta(g)= g\left[-\eps + \frac{g(d-1)C_{d}}{2d}\right].
\label{26}
\end{equation}
From (\ref{26}) it follows that an IR-attractive fixed point
\begin{equation}
 g_* = \frac{2d\eps}{C_{d}\,(d-1)}
\label{260}
\end{equation}
of the RG equations [$\beta (g_*) = 0$, $\beta '(g_*)=\eps > 0$]
exists in the physical region $g>0$ for all $\eps>0$. The value
of $\gamma_{\nu}(g)$ at the fixed point is also found  exactly:
\begin{equation}
\gamma_{\nu}^{*} \equiv \gamma_{\nu}(g_*)= \eps ,
\label{27}
\end{equation}
without corrections of order $\eps^{2}$, $\eps^{3}$, and so on.

It is well known that the leading term of the IR asymptotic
expression of any renormalized quantity $G^{R}$, for which the
RG equation of the form (\ref{22}) is valid, satisfies the same
equation with the substitution $g \to g_*$, where $g_*$ is the
IR stable fixed point:
\begin{equation}
\left[{\cal D}_{\mu}-\gamma_{\nu}^{*}{\cal D}_{\nu}+
\gamma^{*}_{G}\right]G^{R} = 0, \quad
\gamma^{*}_{G}\equiv \gamma^{*}_{G}(g_{*}).
\label{28}
\end{equation}
Canonical scale invariance is expressed by the relations:
\begin{equation}
\left[\sum _{\alpha}d_{\alpha}^k{\cal D}_{\alpha}-
d_{G}^k\right]G^{R}=0 ,\quad
\left[\sum _{\alpha}d_{\alpha}^{\omega }{\cal D}_{\alpha}-
d_{G}^{\omega }\right]G^{R}=0 ,
\label{29}
\end{equation}
in which $\alpha\equiv\{t,{\bf x},\mu,\nu,m,M,g\}$ is the set of
all arguments of $G^{R}$ ($t,{\bf x}$ is the set of all times
and coordinates), and $d^{k}$ and $d^{\omega}$ are the
canonical dimensions of $G^{R}$ and $\alpha$. Substituting the
needed dimensions from Table \ref{table1} into (\ref{29}), we obtain:
\begin{eqnarray}
\left[{\cal D}_{\mu}+{\cal D}_{m}+{\cal D}_{M}-2{\cal D}_{\nu}
-{\cal D}_{\bf x}-d_{G}^{k}\right]G^{R} = 0,\nonumber \\
\left[{\cal D}_{\nu}-{\cal D}_{t}-d_{G}^{\omega}\right]G^{R} = 0.
\label{30}
\end{eqnarray}

Each of the equations (\ref{28})--(\ref{30}) describes scaling
with dilatation of all variables, the derivatives with respect
to which enter into the differential operator. We are interested
in scaling with dilatation of $t,{\bf x}$, and ``masses'' $M,m$
for fixed $\mu,\nu$, and $ g$, and it is necessary to exclude
the corresponding derivatives ${\cal D}_{\alpha}$ by a combination
of the available equations. After eliminating ${\cal D}_{\mu}$
and ${\cal D}_{\nu}$  from (\ref{28}) and (\ref{30}) we obtain
the desired equation of critical IR scaling for the model
(\ref{10}):
\begin{equation}
\left[-{\cal D}_{\bf x}+ \Delta_{t} {\cal D}_{t} +
\Delta_{m} {\cal D}_{m} + \Delta_{M} {\cal D}_{M} -
\Delta_{G} \right]G^{R} = 0
\label{31}
\end{equation}
with the coefficients
\begin{equation}
\Delta_{t}=-\Delta_{\omega}=-2+\gamma^{*}_{\nu}=-2+\eps,\quad
\Delta_{m} = \Delta_{M} =1
\label{32A}
\end{equation}
and
\begin{equation}
\Delta[G]\equiv\Delta_{G} = d_{G}^{k}+ \Delta_{\omega}
d_{G}^{\omega}+\gamma_{G}^{*},
\label{32B}
\end{equation}
which are the corresponding critical dimensions.
In particular, for any correlation function
$G^{R}=W^{R}=\langle \Phi\dots\Phi\rangle$ of the fields $\Phi$
we have $\Delta_{G} = N_{\Phi} \Delta_{\Phi}$, with the summation
over all fields $\Phi$ entering into $ G^{R}$, and for
$\Delta_{\Phi}$ using the data from Table \ref{table1} and the exact value
of $\gamma^{*}_{\nu}= \eps$ we obtain from (\ref{32B}) the following
exact expressions:
\begin{equation}
\Delta_{\bf v}=1-\eps,\quad
\Delta_{\theta} = -1+\eps/2,\quad
\Delta_{\theta'} = d+1-\eps/2
\label{33}
\end{equation}
(we recall that the fields in the model (\ref{10}) are
not renormalized and therefore $\gamma_{\Phi}=0$ for all $\Phi$).

To avoid misunderstanding, we again emphasize the fact that the
RG equation (\ref{31}) describes  IR scaling, i.e., the statement
which is equivalent to the critical scaling in the theory of
critical phenomena, see also discussion in \cite{UFN,JETP}.
In this  scaling the variables $M$, $m$ are also IR
relevant, i.e., they are dilated in scale transformations.
In other words, the expression (\ref{31}) describes the
asymptotic behaviour as $\Lambda r>>1$ for any fixed
values of $Mr$ and $mr$, with  the UV scale  $\Lambda$
defined in (\ref{g0}). The solution of the set of equations
(\ref{30}), (\ref{31}) can be found only up to some unknown
function of all the first integrals, including those of the
form $mr, Mr$, where $r=|{\bf x}_{i}-{\bf x}_{j}|$ is some
coordinate difference.  The anomalous exponents
in the expressions (\ref{HZ1}), (\ref{HZ2}) describe
the behaviour of the corresponding correlation functions for
$Mr\to0$, and therefore are {\it not} related to the their
critical dimensions (\ref{32B}).

In terms of unrenormalized variables, the solution
of the set of equations (\ref{29}), (\ref{31})
for the example of a single-time pair correlation function
can be found, apart from numerical factor, in the form
(for more details, see e.g. \cite{UFN}):
\begin{equation}
G\propto G^{R}\simeq  D_{0}^{d_{G}^{\omega}}\,
\Lambda^{-\gamma_{G}^{*}}\, r^{-\Delta_{G}}\, f(Mr,mr),
\label{1.55}
\end{equation}
with certain, as yet unknown, scaling function $f$.

In what follows we limit ourselves to the correlation
functions of the form (\ref{HZ1}), (\ref{HZ2}),
which are finite for $m=0$, see [31--34].
We shall always set $m=0$ and study the dependence of
the scaling functions $f(Mr)\equiv f(Mr,mr=0)$ on the only
remaining argument $Mr$  in the asymptotic region $Mr<<1$.
This can be performed within the framework of the general
solution (\ref{1.55}) of the RG equations with the aid of
additional methods, see Sec.\ref{sec-4}.

\section{Renormalization and critical dimensions of
com\-po\-si\-te operators} \label {sec-2}

The quantities entering into the left hand sides of
Eqs.(\ref{HZ1}), (\ref{HZ2}) are two-point correlation functions
of composite fields (composite operators in quantum-field
terminology) rather than multipoint correlators of the primary
fields. In what follows, we use the term ``composite operator''
for any local (unless stated to be otherwise) monomial or polynomial
constructed from primary fields and their derivatives at a
single point $x\equiv (t,{\bf x})$. Examples are $\theta^{n}(x)$,
$[\partial_{i}\theta(x)\partial_{i}\theta(x)]^{n}$,
$\partial_{i}\theta(x)\partial_{j}\theta(x)$,
$\theta'(x)\nabla_{t}\theta(x)$ and so on.

Since the arguments of the fields coincide, correlation functions with
these operators contain additional UV divergences, which are removed
by additional renormalization procedure, see
e.g. \cite{Collins}. For the renorma\-li\-zed correlation functions
standard RG equations are obtained, which describe IR scaling with
definite critical dimensions $\Delta_{F}\equiv\Delta[F]$  of
certain ``basis'' operators  $F$. Owing to the renormalization,
$\Delta[F]$ does not coincide in general with the na\"{\i}ve sum
of critical dimensions of the fields and derivatives entering
into $F$.

Detailed exposition of the re\-nor\-ma\-li\-za\-ti\-on of composite
operators for the stochastic Navier--Stokes equation is given
in the review paper \cite{UFN}, see also [13--17,19--21],
below we confine ourselves to only the necessary information.

In general, composite operators are mixed in renormalization,
i.e., an UV finite renormalized operator $F^{R}$ has the form
$F^{R}=F+$ counterterms, where the contribution of the
counterterms is a linear combination of $F$ itself and,
possibly, other unrenormalized operators which ``admix''
to $F$ in renormalization.

Let $F\equiv\{F_{\alpha}\}$ be a closed set, all of whose
monomials mix only with each other in renormalization.
The renormalization matrix $Z_{F}\equiv\{Z_{\alpha\beta}\}$
and the  matrix of anomalous dimensions
$\gamma_{F}\equiv\{\gamma_{\alpha\beta}\}$
for this set are given by
\begin{equation}
F_{\alpha }=\sum _{\beta} Z_{\alpha\beta}
F_{\beta }^{R},\qquad
\gamma _F=Z_{F}^{-1}\D_{\mu }Z_{F},
\label{2.2}
\end{equation}
and the corresponding matrix of critical dimensions
$\Delta_{F}\equiv\{\Delta_{\alpha\beta}\}$ is given by Eq.(\ref{32B}),
in which
$d_{F}^{k}$, $d_{F}^{\omega}$, and $d_{F}$ are understood as the
diagonal matrices of canonical dimensions of the operators in
question (with the diagonal elements equal to sums of corresponding
dimensions of all fields and derivatives constituting $F$) and
$\gamma^{*}\equiv\gamma (g_{*})$ is the matrix (\ref{2.2}) at
the fixed point.

Critical dimensions of the set $F\equiv\{F_{\alpha}\}$ are
given by the eigenvalues of the matrix $\Delta_{F}$. The ``basis''
operators that possess definite critical dimensions have the form
\begin{equation}
{\bar F}^{R}_{\alpha}=\sum_{\beta}
U_{\alpha \beta}F^{R}_{\beta}\ ,
\label{2.5}
\end{equation}
where the matrix $ U_{F} =  \{U_{\alpha \beta} \}$
is such that $ \Delta'_{F}= U_{F} \Delta_{F} U_{F}^{-1}$
is diagonal.

In general, counterterms to a given operator $F$ are
determined by all possible 1-irreducible Green functions
with one operator $F$ and arbitrary number of primary fields,
$\Gamma=\langle F(x) \Phi(x_{1})\dots\Phi(x_{2})\rangle$.
The total canonical dimension (formal index of divergence)
for such functions is given by
\begin{equation}
d_\Gamma = d_{F} - N_{\Phi}d_{\Phi},
\label{index}
\end{equation}
with the summation over all types of fields entering into
the function. For superficially divergent diagrams,
$d_\Gamma$ is a nonnegative integer, cf. Sec.\ref{sec-1}.

Let us consider operators of the form $\theta^{n}(x)$
with the canonical dimension $d_{F}=-n$, entering into the
structure functions (\ref{HZ1}). From Table \ref{table1}
in Sec.\ref{sec-1} and Eq.(\ref{index}) we obtain
$d_\Gamma = -n+N_{\theta}-N_{\bf v} -(d+1)N_{\theta'}$,
and from the analysis of the diagrams it follows that the total
number of the fields $\theta$ entering into the function
$\Gamma$ can never exceed the number of the fields $\theta$
in the operator $\theta^{n}$ itself, i.e., $N_{\theta}\le n$.
Therefore, the divergence can only exist in the functions
with $N_{\bf v}=0$, $N_{\theta'}=0$, and arbitrary value of
$n=N_{\theta}$, for which the formal index vanishes, $d_\Gamma =0$.
However, at least one of $N_{\theta}$ external ``tails''
of the field $\theta$ is attached to a vertex
$\theta'({\bf v}\partt)\theta$ (it is impossible to construct
nontrivial, superficially divergent diagram of the desired
type with all the external tails attached to the vertex $F$),
at least one derivative $\partial$ appears as an extra
factor in the diagram, and, consequently, the real index of
divergence is necessarily negative, see Sec.\ref{sec-1}.

This means that the operator $\theta^{n}$ requires no
counterterms at all, i.e., it  is in fact UV
finite, $\theta^{n}=Z\,[\theta^{n}]^{R}$ with $Z=1$.
It then follows that the
critical dimension of $\theta^{n}(x)$ is simply given by
the expression  (\ref{32B})
with no correction from $\gamma_{F}^{*}$ and is therefore reduced
to the sum of the critical dimensions of the factors:
\begin{equation}
\Delta [\theta^{n}] = n\Delta[\theta] =n (-1+\eps/2).
\label{2.6}
\end{equation}
We note that this relation was not clear {\it a priori} and
it is a specific feature of the model (\ref{10}).
For example, in the standard model $\phi^{4}$ of the theory of
critical phenomena, the critical dimensions of the field
$\phi$ and the composite operators $\phi^{2}$ and $\phi^{4}$
are completely independent, and they determine independent
critical exponents $ \eta, \nu$, and $ \omega$, see e.g.
\cite{Zinn}. The relation analogous to (\ref{2.6}) is valid
for the powers of the velocity field of the stochastic
Navier--Stokes equation, where it is a consequence of the
Galilean symmetry of the model, see e.g. \cite{UFN,JETP,Kim}.

Now let us turn to the operators $F_{n}\equiv [\partial_{i}\theta
\partial_{i}\theta]^{n}$ with $d_{F}=0$, $d_{F}^{\omega}=-n$.
They enter into the left-hand sides of Eqs.(\ref{HZ2}) and,
as we shall see in  Sec.\ref{sec-4}, it is their critical
dimensions that determine the anomalous exponents in
(\ref{HZ1}) and (\ref{HZ2}).

In this case, from Table \ref{table1} in Sec.\ref{sec-1}
and Eq.(\ref{index}) we have
$d_\Gamma = N_{\theta}-N_{\bf v} -(d+1)N_{\theta'}$,
with the necessary condition $N_{\theta}\le 2n$, which follows
from the structure of the diagrams. It is also obvious from the
analysis of the diagrams that the counterterms to these operators
can involve the fields $\theta$, $\theta'$ only in the form of
derivatives, $\partial\theta$, $\partial\theta'$,
and so the real index of divergence
has the form $d_\Gamma' = d_\Gamma -N_{\theta}-N_{\theta'}=
-N_{\bf v} -(d+2)N_{\theta'}$.   It then follows that
superficial divergences exist only in the Green functions with
$N_{\bf v}=N_{\theta'}=0$  and any $N_{\theta}\le 2n$,
and the corresponding operator counterterms are reduced to the
form $F_{k}$ with $k\le n$. Therefore, the operators
$F_{n}$ can mix only with each other in renormalization,
the corresponding infinite renormalization matrix
$Z_{F}=\{Z_{nk}\}$ is in fact triangular, $Z_{nk}=0$ for $k>n$,
and the critical dimensions associated with the operators $F_{n}$
are determined by the diagonal elements $Z_{n}\equiv Z_{nn}$
(in contradistinction with the case of operators  $\theta^{n}$,
they are not equal to unity here).

Explicit calculation of the constants $Z_{n}$ in the MS scheme
in the two-loop approximation gives:
\begin{eqnarray}
Z_{n}^{-1}= 1- \frac{u}{2\eps}\cdot \frac{n(d-1)(d+2n)}{d(d+2)}+
\nonumber \\
+\frac{u^{2}}{8\eps^{2}}\cdot \frac{n(n-1)(d-1)^{2}(d+2n)(d+2n+2)}
{d^{2}(d+2)^{2}}+  \nonumber \\
+\frac{u^{2}}{2\eps} \cdot \frac{n(n-1)(d-1)}
{d^{2}(d+2)^{2}(d+4)} \biggl[ \frac{-(d+4)(d+1)}{(d+2)} +
\nonumber \\
+ {3(d-1)(d+2n)}
h_{1}(d)/4-\frac{(d+1)(d+3n-2)}{(d+4)}
h_{2}(d)\biggr] + O(u^{3}),
\label{Zn}
\end{eqnarray}
where we have written $u\equiv gC_{d} $
with the coefficient $C_{d}$ from (\ref{25}), and
\begin{eqnarray}
h_{1}(d)\equiv F(1,1,d/2+2,1/4), \nonumber \\
h_{2}(d)\equiv F(1,1,d/2+3,1/4)= h_{1}(d+2),
\label{u}
\end{eqnarray}
where $F(\dots)$
is the hypergeometric series (see e.g. \cite{Gamma}):
\begin{equation}
F(a,b,c,z)\equiv 1+\frac{ab}{c}\, z+ \frac{a(a+1)b(b+1)}{c(c+1)}
\cdot \frac{z^{2}}{2!}+\dots.
\label{hyper}
\end{equation}
From (\ref{u})  and (\ref{hyper})  it follows:
\begin{equation}
h_{1}(d)=\sum_{k=0}^{\infty}\frac{k!\, \Gamma(d/2+2)}
{4^{k}  \Gamma(d/2+2+k)},\quad h_{2}(d)=h_{1}(d+2).
\label{hyper2}
\end{equation}
Representations (\ref{hyper2}) are convenient for numerical
computations of the functions $h_{1,2}(d)$.

From (\ref{Zn}) for the anomalous dimension $\gamma _n\equiv
\D_\mu \ln Z_{n}$ in the two-loop order we obtain:
\begin{eqnarray}
\gamma _n (g)=-\frac{un(d-1)(d+2n)}{2d(d+2)}+ \frac{u^{2}n(n-1)}
{d(d+2)(d+4)}\times
\nonumber \\
\times\left[ \frac{(d+4)(1-d^{2})}{d(d+2)^{2}}
+\frac{3(d-1)^{2}(d+2n)}
{4d(d+2)}h_{1}(d)-\frac{(d^{2}-1)(d+3n-2)}{d(d+2)(d+4)}
h_{2}(d)\right] + O(u^{3}),
\label{Gn}
\end{eqnarray}
and for the corresponding critical dimension $\Delta_{n}$ at the
fixed point (\ref{260})
using (\ref{32B}) and $d_{F}[F_{n}]=0$, $d_{F}^{\omega}[F_{n}]=-n$
(see Table \ref{table1} in Sec.\ref{sec-1}) we have:
\begin{eqnarray}
\Delta_n= n\eps+ \gamma_{n}^{*}= -\frac{2n(n-1)\eps}{d+2}+
\frac{\eps^{2}n(n-1)}{(d-1)(d+2)^{3}(d+4)^{2}}\times
\nonumber \\
\times\biggl[-4(d+1)(d+4)^{2}+3(d-1)(d+2)(d+4)(d+2n)h_{1}(d)-
\nonumber \\
-  4(d+1)(d+2)(d+3n-2)h_{2}(d) \biggr]+ O(\eps^{3}).
\label{Dn}
\end{eqnarray}

Expression (\ref{Dn}) is simplified for any integer value
of $d$ owing to the fact that the series in (\ref{hyper})
reduce to finite sums, see \cite{Gamma}:
\begin{equation}
h_{1}(d)=2(d+2)\left[(-3)^{d/2}\ln(4/3) +
\sum_{k=2}^{d/2+1} \frac{(-3)^{k-2}} {d/2-k+2}\right]
\label{Gamma4}
\end{equation}
for any even value of $d$ and
\begin{equation}
h_{1}(d)=2(d+2)\left[(-1)^{(d+1)/2}\cdot 3^{d/2-1}\cdot\pi+2\,
\sum_{k=1}^{(d+1)/2} \frac{(-3)^{(d+1)/2-k}} {2k-1}\right]
\label{Gamma3}
\end{equation}
for any odd value of  $d$, which for $d=2$  and $d=3$ gives
the results announced in (\ref{d=2}), (\ref{d=3}).

We explain in Sec.\ref{sec-4} that the critical
dimensions $\Delta_{n}$ from (\ref{Dn}) are nothing else than the
anomalous exponents entering into relations (\ref{HZ1}),
(\ref{HZ2}), and
here we only note that the first order in $\eps$ of the
expression (\ref{Dn}) coincides (up to the notation) with
the result obtained in \cite{GK} for $n=2$ and in
\cite{BGK} for arbitrary $n$, and that the first term of the
expansion in $1/d$ of the expression (\ref{Dn}) coincides with
the result (\ref{HZ3}) obtained in \cite{Falk1,Falk2}.
We also note that the $\eps^{2}$ term of the expression (\ref{Dn})
behaves as  $1/d^{2}$  for $1/d\to0$ and therefore gives no
contribution to the first order of the $1/d$ expansion
[this follows from the relation $h_{2}(d)= h_{1}(d+2)=1+O(1/d)$,
which is obvious from (\ref{hyper2})]. This fact
suggests that the $\eps$ expansion for the dimension $\Delta_n$
in the model (\ref{10}) is ``better'' than the $1/d$ expansion
in the sense that a given order of the $1/d$ expansion is
contained completely in the corresponding order of the
$\eps$  expansion, but not the reverse. We also note that
the $n$ dependence of the quantity $\Delta_{n}$ in the second
order of the $\eps$ expansion is no longer reduced to
the simple factor $n(n-1)$.

The result $\Delta_{1}=0$ in (\ref{Dn}) is in fact exact,
in agreement with the exact solution for the
two-point structure function obtained in \cite{Kraich1}.
Within the RG approach this can be demonstrated using the
Schwinger equation of the form
\begin{equation}
\int{\cal D}\Phi {\delta} \left[ \theta(x) \exp S_{R}( \Phi)
+ A \Phi\right]/{\delta\theta'(x)}  =0
\label{Schwi}
\end{equation}
(in the general sense of the term, Schwinger equations are any
relations stating that any functional integral of a total
variational derivative is equal to zero, see e.g. Sec.7 of
\cite{Book}). In (\ref{Schwi}),  $S_{R}$ is the renormalized analog
of the action (\ref{10}), and the notation introduced in (\ref{14})
is used.  Eq.(\ref{Schwi}) can be rewritten in the form
\begin{equation}
\langle\langle \theta' D_{\theta} \theta - \nabla_{t}[\theta^{2}/2] +
\nu Z_{\nu}\triangle[\theta^{2}/2]-\nu Z_{\nu} F_{1}\rangle\rangle
_{A}=-A_{\theta'} \delta W_{R}(A)/\delta A_{\theta}.
\label{Schwi2}
\end{equation}
Here $D_{\theta}$ is the correlator (\ref{2}),
$\langle\langle \dots\rangle\rangle _{A}$
denotes the averaging with the weight $ \exp [S_{R}( \Phi) +
A \Phi]$, $W_{R}$ is determined by the Eq.(\ref{14}) with
the replacement $S\to S_{R}$, and the argument $x$ common to all
the quantities in (\ref{Schwi2}) is omitted.

The quantity $\langle\langle F\rangle\rangle _{A}$ is the
generating functional of the
correlation functions with one operator $F$ and any number of
fields $\Phi$, therefore the UV finiteness of the operator $F$ is
equivalent to the finiteness of the functional
$\langle\langle F\rangle\rangle _{A}$.
The quantity in the right hand side of Eq.(\ref{Schwi2}) is
finite (a derivative of the renormalized functional
with respect to finite argument), and so is the operator in the
left hand side. Our operator $F_{1}$  does not admix in
renormalization to the operator $\theta' D_{\theta}\theta$
($F_{1}$ contains too many fields $\theta$), and to the
operators $\nabla_{t}[\theta^{2}/2]$ and $\triangle[\theta^{2}/2]$
(they have the form of total derivatives, and $F_{1}$ is not
reduced to this form). On the other hand, the operator
$\theta' D_{\theta}\theta$ does not admix to $F_{1}$ (it is
nonlocal, and $F_{1}$ is local), while the derivatives
$\nabla_{t}[\theta^{2}/2]$ and $\triangle[\theta^{2}/2]$
do not admix to $F_{1}$ owing to the fact that each field
$\theta $ enters in the counterterms of the operators
$F_{n}$ only in the form of derivative $\partial\theta$
(see above). Therefore, all three types of operators entering
into the left hand side of the Eq.(\ref{Schwi2}) are independent,
and they must be UV finite separately.

Since the operator $\nu Z_{\nu} F_{1}$ is UV finite, it coincides
with its finite part, i.e., $\nu Z_{\nu}F_{1}=\nu F_{1}^{R}$,
which along with the relation $F_{1}=Z_{1}F_{1}^{R}$ gives
$Z_{1}=Z_{\nu}^{-1}$ and therefore $\gamma_{1}=-\gamma_{\nu}$.
For the critical exponent $\Delta_{1}=\eps + \gamma_{1}^{*}$
we then obtain $\Delta_{1}=0$ exactly (we recall that
$\gamma_{\nu}^{*}=\eps$, see Sec.\ref{sec-1}).

In the SNS model, critical dimensions of certain composite
operators can sometimes be obtained exactly using various
Schwinger equations and Ward identities for Galilean
transformations, see  [12--17] and [19--21].
In particular, exact critical
dimension of the energy dissipation rate $E(x)$ was found in
\cite{Pismak}, see also \cite{Tensor}. The simple relation
$\Delta[E^{n}]=n \Delta[E]$ for the powers of the dissipation
rate was proposed in \cite{Yakhot1}. It was explained later in
\cite{Eight}, that this relation can not be considered reliable,
and here we only note that even for the simple Obukhov--Kraichnan
model, the critical dimension of $E^{n}$ is not a linear function
in $n$, see (\ref{Dn}), in contradistinction with the
powers of the field itself, see (\ref{2.6}) and
\cite{UFN,JETP,Kim} for the case of the SNS model.

Critical dimensions of various tensor operators of the form
$\partial_{i_{1}}\theta(x)\dots\partial_{i_{n}}\theta(x)$
can also be calculated in the second order of the $\eps$
expansion from the same two-loop diagrams, which
determine the constants (\ref{Zn}); only the symmetry
coefficients differ from those for scalar operators.
We shall confine ourselves to the second-rank irreducible
traceless tensors of the form
\begin{equation}
F_{ij}^{n}\equiv\partial_i\theta\,\partial_j\theta\, F_{n-1}-
\delta_{ij}\, F_{n}/d,
\label{Tens}
\end{equation}
where $ F_{n}$ are the scalar operators discussed above. The
operators (\ref{Tens}) mix only with each other in renormalization,
the corresponding renormalization matrix is triangular and its
diagonal elements determine the corresponding critical dimensions
$\Delta'_{n}$. In the second order of the $\eps$ expansion we
have obtained:
\begin{eqnarray}
\Delta'_n= \frac{\eps[d(d+1)-2n(n-1)(d-1)]}{(d-1)(d+2)}+
\frac{\eps^{2}n(n-1)}{(d-1)(d+2)^{3}(d+4)^{2}}\times
\nonumber \\
\times\biggl[-4(d+1)(d+4)^{2}P_{0}+
3(d-1)(d+2)(d+4)
(d+2n)h_{1}(d)P_{1}-
\nonumber \\
-4(d+1)(d+2)(d+3n-2)h_{2}(d)P_{2} \biggr]+ O(\eps^{3}),
\label{Tensor2}
\end{eqnarray}
where we have written:
\begin{eqnarray}
P_{0}=1+\frac{d}{2n(n-1)(d-1)},\nonumber \\
P_{1}=1-\frac{d(d+1)}{n(d-1)(d+2n)},
\nonumber  \\
P_{2}=1-\frac{d(2d-1)}{2n(d-1)(d+3n-2)},
\label{Polinom}
\end{eqnarray}
and the functions $h_{1,2}(d)$ are defined in (\ref{u}).

We note that the expression (\ref{Tensor2}) for $n=1$ coincides
with the first two terms of the  expansion in $\eps$ of the
exact result obtained in \cite{Falk1} (we explain below
that the quantity $\Delta'_1$ corresponds to $\delta-\gamma $
in the notation of \cite{Falk1}). For the case of the SNS model,
critical dimensions of irreducible tensor operators are
studied, for example, in \cite{Triple}.

Since the critical dimensions of the operators $\theta^{n}$
and $(\partial\theta)^{n}$ have been found, we can
use the general expression (\ref{1.55}) in the case of
the structure functions (\ref{HZ1}) and pair correlators
(\ref{HZ2}).

The structure function $S_{2n}(r)=
\langle[\theta(x)-\theta(x')]^{2n} \rangle$ is represented
as a sum of pair correlators
$\langle\theta(x)^{k}\theta(x')^{m}\rangle$
with fixed value of
$k+m=2n$ and equal canonical and critical dimensions
$d_{F}^{\omega}=2nd_{\theta}^{\omega}=-n$,
$\Delta_{F}=2n\Delta_{\theta}=n(-2+\eps)$.
Then from Eq.(\ref{1.55}) with $m=0$ it follows:
\begin{equation}
S_{2n}(r) \simeq D_{0}^{-n}\, r^{n(2-\eps)}\, f_{n}(Mr),
\label{100}
\end{equation}
with certain, as yet unknown, scaling functions $f_{n}(Mr)$.

The $n$th power of the dissipation rate is represented as
a finite linear combination of basis operators (\ref{2.5})
with definite critical dimensions
$\Delta_{k}=k\eps+\gamma^{*}_{k}$
given in (\ref{Dn}), with the necessary condition $k\le n$.
Therefore, for the pair correlators (\ref{HZ2}) of the powers
of the dissipation rate we obtain from (\ref{1.55}):
\begin{equation}
\langle E^{n}(x)E^{m}(x')\rangle = \nu_0 ^{n+m}
\langle F_{n}(x)F_{m}(x')\rangle\simeq
\sum_{k=0}^{n} \sum_{l=0}^{m}
(\Lambda r)^{-\Delta_{k}-\Delta_{l}}  f_{k,l}\,(Mr),
\label{1010}
\end{equation}
with certain scaling functions $ f_{k,l}\,(Mr)$ and the
UV scale $\Lambda$ defined in (\ref{g0}). The leading term
of the asymptotic
behaviour of the expression  (\ref{1010}) in the IR region
$\Lambda r>>1$ is given by the contribution with minimal
$\Delta_{k}+\Delta_{l}$, i.e.,
\begin{equation}
\langle E^{n}(x)E^{m}(x')\rangle  \simeq
(\Lambda r)^{-\Delta_{n}-\Delta_{m}} f_{n,m}(Mr).
\label{101}
\end{equation}

We note that the structure functions (\ref{100}) and the correlator
(\ref{101}) for $n=m=1$ are independent of the diffusivity
coefficient, or, equivalently, of the UV scale $\Lambda$ (we recall
that $\Delta_{1}=0$). This statement is an
analog of the Second Kolmogorov hypothesis (independence of the
viscosity coefficient in the inertial and energy-containing ranges)
for the real fully developed turbulence, see e.g. \cite{Monin}
and \cite{Orszag}.  Within the RG approach to the
SNS model, the Second Kolmogorov hypothesis was established in
\cite{JETP}, see also \cite{UFN,AV}.

Expressions like (\ref{100}), (\ref{101})  can easily be written
down for any single-time pair correlator, provided its canonical
and critical dimensions are known. Let us give two more examples.
For the second-rank irreducible tensors $E'_{n}=\nu_0 F'_{n}$ with
the operators $F'_{n}$ given in (\ref{Tens}) we obtain
(dropping the vector indices):
\begin{equation}
\langle E'_{n}(x)E'_{m}(x')\rangle  = \nu_0 ^{n+m}
\langle F'_{n}(x)F'_{m}(x')\rangle\simeq
(\Lambda r)^{-\Delta'_{n}-\Delta'_{m}}  f''_{n,m}(Mr),
\label{102}
\end{equation}
and for the mixed correlator of the scalar  $E^{n}$ and
the tensor $E_{n}'$  we have:
\begin{equation}
\langle E^{n}(x)E'_{m}(x')\rangle  = \nu_0 ^{n+m}
\langle F_{n}(x)F'_{m}(x')\rangle\simeq
(\Lambda r)^{-\Delta_{n}-\Delta'_{m}}  f'_{n,m}(Mr),
\label{103}
\end{equation}
with the dimensions $\Delta_{n}$, $\Delta'_{n}$ given in
(\ref{Dn}), (\ref{Tensor2})
 and certain scaling functions $f'_{n,m}(Mr)$,
$f''_{n,m}(Mr)$ (the prime is not a derivative here).

\section{Operator product expansion and anomalous scaling}
\label {sec-4}

From the viewpoint of the renormalization group,
the expressions (\ref{1.55}) and (\ref{100})--(\ref{103})
for any functions $f(Mr)$
correspond to IR scaling in the region $L\equiv M^{-1}$,
$r>>l\equiv \Lambda^{-1}$ for arbitrary fixed value of $Mr$,
with definite critical dimensions $\Delta_{F}$. The inertial
range $l<<r<<L$ corresponds to the additional condition $Mr<<1$,
and representations like (\ref{HZ1}) and (\ref{HZ2}) should be
understood as certain additional statements about the explicit
form of the leading terms of the asymptotic behaviour for $Mr\to0$.

In the theory of critical phenomena, the asymptotic form of
 scaling functions for  $M\to0$ is studied using the well
known Wilson operator product expansion (OPE), see e.g.
 \cite{Collins,Zinn}; the analog of $L\equiv M^{-1}$
is there the correlation length $r_{c}$.
This technique is also applied to the theory of
turbulence, see e.g. \cite{UFN,JETP}.

According to the OPE, the single-time product
$F_{1}(x_{1})F_{2}(x_{2})$
of two renormalized operators at
${\bf x}\equiv ({\bf x}_{1} + {\bf x}_{2} )/2 = {\const}$, and
${\bf r}\equiv {\bf x}_{1} - {\bf x}_{2}\to 0$
has the representation
\begin{equation}
F_{1}(x_{1})F_{2}(x_{2})=\sum_{\alpha}C_{\alpha} ({\bf r})
\bar F^{R}_{\alpha}({\bf x,t}) ,
\label{2.44}
\end{equation}
in which the functions $C_{\alpha}$  are the Wilson coefficients
regular in $M^{2}$ and  $\bar F^{R}_{\alpha}$ are all possible
renormalized local composite operators of the type (\ref{2.5})
allowed by symmetry, with definite critical
dimensions $\Delta_{\alpha}$.

The renormalized correlator $\langle F_{1}(x_{1})F_{2}(x_{2})
\rangle$ is obtained by averaging (\ref{2.44}) with the weight
$\exp S_{R}$, the quantities  $\langle\bar F^{R}_{\alpha}\rangle
=M^{d_{\alpha}}\nu^{d_{\alpha}^{\omega}} a_{\alpha} (g,M/\mu)$
involving dimensionless functions $ a_{\alpha} (g,M/\mu)$
appear on the right hand side. Their asymptotic behaviour
for $M/\mu\to0$ is found from the corresponding RG equations and
has the form
\begin{equation}
\langle\bar F^{R}_{\alpha}\rangle \propto  M^{\Delta_{\alpha}}.
\label{2.45}
\end{equation}
From the operator product expansion (\ref{2.44}) we therefore
find the following expression  for the scaling function
$f(Mr)$ in the representation  (\ref{1.55})  for the correlator
$\langle F_{1}(x_{1})F_{2}(x_{2})\rangle$:
\begin{equation}
f(u)=\sum_{\alpha}A_{\alpha}(u)\, u^{\Delta_{\alpha}},\quad
u\equiv Mr,
\label{2.46}
\end{equation}
with coefficients $A_{\alpha}$, which are regular in  $u^{2}$,
generated by the Wilson coefficients $C_{\alpha}$ in (\ref{2.44}),
which are regular in $M^{2}$.

The leading contributions for  $u\to0$
are those with the smallest dimension $\Delta_{\alpha}$
and in the $\eps$ expansions they are those with the smallest
$d_{\alpha}\equiv d[F_{\alpha}]$  for $\eps=0$.
We shall term the operators with $\Delta_{\alpha}<0$,
if they exist, dangerous \cite{UFN,JETP},
as they correspond to contributions to (\ref{2.46})
which diverge for $u\to0$.

In the standard model $\phi^{4}$ of the theory of critical
behaviour \cite{Zinn}, there are no problem of dangerous
operators within the $\eps$ expansions, because in that model
$\Delta_{\alpha} = n_{\alpha} + O(\eps)$,
where $n_{\alpha}\ge0$  is the total number of fields and
derivatives in $F^{R}_{\alpha}$.
The operator $F=1$  has the smallest value $n_{\alpha}=0$,
but it gives a contribution to (\ref{2.46})  which is regular
in  $u^{2}$ and has a finite limit as $u\to0$.
The first nontrivial contribution is generated by the operator
$\phi^{2}$ with $n_{\alpha}=2$, it has the form  $u^{2+O(\eps)}$
and only determines correction to the leading term
generated by the operator $F=1$, which vanishes at $u\to0$.

We note that for a Galilean invariant product $F_{1}(x_{1})F_{2}
(x_{2})$, the right hand side of Eq.(\ref{2.44})
can involve any Galilean invariant operator, including tensor
operators, whose indices would be contracted with the analogous
indices of the coefficients $C_{\alpha}$. Without loss of
generality, it can be assumed that the expansion is made in
irreducible tensors (see Sec.\ref{sec-2} for examples),
so that only scalars contribute to the correlator
$\langle F_{1}F_{2}\rangle$  because the averages
$\langle\bar F^{R}_{\alpha}\rangle$  for nonscalar irreducible
tensors are zero. For the same reason, the contributions to the
correlator from all operators of the form $\partial F$  with
external derivatives vanish owing to translational invariance.

In our case, contributions from the operators like $\theta^{n}$
with negative $d_{F}$ are also forbidden by the invariance
of the correlators (\ref{100})--(\ref{103}) with respect to the
shift  $\theta(x)\to\theta(x)+\const$ of the field $\theta$.

The leading terms of the asymptotic behaviour of the scaling
functions in (\ref{100})--(\ref{103})  for $Mr\to0$
are therefore determined by the  scalar operators
$ F_{n}=[\partial_{i}\theta\partial_{i}\theta]^{n}$
with the minimal canonical dimension $d_{F}=0$, see
Sec.\ref{sec-2}. From the analysis of the diagrams it follows
that the number of the fields $\theta$ in the operator $F_{n}$
entering into the right hand sides of the expansions (\ref{2.44})
can never exceed the total number of the  fields $\theta$
in their left hand sides. Therefore, only finite number of
operators $ F_{n}$ contributes to each operator product
expansion in the model (\ref{10}), and the
asymptotic form  of the scaling functions in
(\ref{100})--(\ref{103}) for $u\to0$ is given by the expression:
\begin{equation}
f(u)=\sum_{n=0}^{N} C_{n} \, u^{\Delta_{n}} +\dots,
\label{2.460}
\end{equation}
where $N$  is the total number of the fields $\theta$ in the
left hand sides,  $C_{n}=C_{n}(\eps, d)$ are numerical
coefficients, and the dots stand for corrections of order
$u^{2+O(\eps)}$ vanishing as  $u\to0$. The leading
term  for $u\to0$ is determined by the operator with minimal
$\Delta_{n}$, i.e., with maximal $n$ (owing to the fact that
the dimension $\Delta_{n}$ within the $\eps$ expansion
decreases monotonically with $n$,
see (\ref{Dn})). As a result, for the scaling
function $f_n$ in the representation (\ref{100}) for the
structure function $S_{2n}$ we obtain:
\begin{equation}
f_{n}\simeq \const u^{\Delta_{n}} \qquad {\rm for}\ u\to0,
\label{201}
\end{equation}
and for the correlators (\ref{101})--(\ref{103}) we have:
\begin{equation}
f_{n,m},\ f'_{n,m},\ f''_{n,m}\simeq \const u^{\Delta_{n+m}}
\qquad {\rm for}\ u\to0,
\label{202}
\end{equation}
with the critical dimensions $\Delta_{n}$  given in (\ref{Dn}).

Therefore, we have derived expressions (\ref{HZ1}), (\ref{HZ2}),
and relate the corresponding anomalous exponents to the critical
dimensions of the composite operators $E^{n}(x)$.

\section{Discussion and conclusion}
\label {sec-5}

We have shown that the re\-nor\-ma\-li\-za\-ti\-on group combined
with the operator product expansion establishes the existence
of anomalous scaling in the model (\ref{1})--(\ref{3})
for the advection of a passive scalar by a Gaussian velocity
field and allows the corresponding anomalous exponents to be
calculated in the form of series in $\eps$.

The distinguishing feature of the model (\ref{1})--(\ref{3})
that explains the origin of the anomalous scaling is
the existence of dangerous composite operators with
{\it negative} critical dimensions. They dominate the asymptotic
behaviour of the scaling functions  and lead to singular
dependence of the correlation functions on the IR scale
$M$ for $M\to0$, in contrast to the standard models
of critical phenomena, in which all the nontrivial operators
have positive critical
dimensions and only determine vanishing corrections to the
leading finite contribution from the simplest operator
$F=1$, see e. g. \cite{Zinn}.
In contradistinction with the SNS model, the dangerous
operators in the model (\ref{1})--(\ref{3}) occur already for
asymptotically small values of $\eps$ and only finite number
of these operators contribute to operator product expansion for
any given correlation function.

The set of expressions (\ref{100})--(\ref{103}),
(\ref{2.460})--(\ref{202}) gives the complete description
of the IR asymptotic behaviour of the Green functions
in the model (\ref{1})--(\ref{3}):
Eqs.(\ref{100})--(\ref{103}) describe the asymptotic form of the
structure functions and pair correlators in the IR region
$L, r >> l\equiv \Lambda$ and determine their
dependence on the UV scale $\Lambda$, while
Eqs.(\ref{2.460})--(\ref{202})  give the asymptotic form of
the corresponding scaling
functions upon additional restriction $Mr<<1$ and determine the
dependence on the IR scale $L\equiv M^{-1}$.
All the critical dimensions $\Delta_{n}$, $\Delta'_{n}$ entering
into (\ref{100})--(\ref{103}), (\ref{2.460})--(\ref{202})  have
been calculated in the second order of the $\eps$  expansion, see
(\ref{Dn}), (\ref{Tensor2}), while the critical dimensions of
the  model parameters, primary fields, and their powers have been
found exactly, see (\ref{32A}), (\ref{32B}), (\ref{33}),
and (\ref{2.6}).

It should be stressed that the asymptotic expressions (\ref{201}),
(\ref{202}) result from the fact that the critical
dimensions $\Delta_{n}$ are negative and that $|\Delta_{n}|$
increases monotonically with $n$. This is obviously so within
the $\eps$  expansion, in which the sign and the $n$ dependence
of the dimensions are determined by the first-order terms,
while the higher-order terms are treated as small corrections.
However, for finite values of $\eps$ the higher-order terms
can, in principle, change these features of the dimensions.
Indeed, the $n^{3}$ contribution in the second-order
approximation for $\Delta_{n}$ is positive, see e.g.
(\ref{d=2}), (\ref{d=3}), and so $\Delta_{n}$ becomes positive,
provided $n$ is large enough. Of course, this conclusion is
based on the second-order approximation of the $\eps$  expansion
and is therefore not reliable: higher-order terms of the $\eps$
expansion contain additional powers of $n$, and the correct
analysis of the large $n$ behaviour of the dimensions
$\Delta_{n}$ requires resummation of the $\eps$ series with
the additional condition that $n\eps\simeq1$, but we know of
no model in which such a resummation has been performed.

The comparison of the
Eqs.(\ref{100})--(\ref{103}), (\ref{201}), (\ref{202})
with the corresponding ex\-pres\-si\-ons in [31--34]
shows that
$\Delta_{n}=-\rho_{2n}$ in the notation of \cite{GK},
$\Delta_{2}=-\Delta$, $\Delta'_{1}=\gamma-\delta$  in
the notation of \cite{Falk1}, and $\Delta_{n}=-\Delta_{2n}$
in the notation of \cite{Falk2};
our results for $\Delta_{n}$, $\Delta'_{n}$ are in
agreement with the results obtained in [31--34]
for the structure functions $S_{2n}(r)$  and the pair
correlators (\ref{101}), and in \cite{Falk1} for the
correlators (\ref{102}), (\ref{103}) with $n=m=1$
within the first order of the expansions in $\eps$ and $1/d$.

It is noteworthy that the set of scalar operators
$F_{n}=[\partial_{i}\theta\partial_{i}\theta]^{n}$
is ``closed with respect to the fusion'' in the sense that
the leading term in the OPE for the pair correlator
$ \langle F_{n} F_{m} \rangle$ is given by the operator
$ F_{n+m} $ from the same family with the summed
index $n+m$, see (\ref{101}), (\ref{202}).
This fact along with the inequality
$\Delta_{n}+\Delta_{m} >  \Delta_{n+m}$, which is obvious
from the explicit expressions for  $\Delta_{n}$,
can be interpreted as the statement that the correlations
of the local dissipation rate in the model
(\ref{1})--(\ref{3}) exhibit multifractal behaviour,
see \cite{DL} and \cite{Eyink1}.  We note that the same
relation ensures the  fulfillment
of the H\"{o}lder inequality  for the structure functions
(\ref{HZ1}).

An important question is that of the universality of
anomalous exponents, see e.g. \cite{ShS}. It is clear
from the RG analysis, that the exponents $\Delta_{n}$ do
not depend on the choice of the correlator (\ref{3}) (this
correlator does not enter at all into the UV divergent
diagrams which determine renormalization constants), and
that they are insensitive to the specific form of the IR
regularization in the correlator (\ref{2}) (renormalization
constants do not depend on the choice of the IR regularization,
see e.g. \cite{Collins}). However, the anomalous exponents can
change if the function $\delta(t-t')$ in the correlator
of the velocity field is replaced by some function with
finite width, i.e., the velocity field has small but finite
correlation time \cite{Falk3}.  The RG approach to this
problem will be presented elsewhere, and here we only
mention another possible modification of the model
(\ref{1})--(\ref{3}), that of a ``slow'' velocity field.
In this case, the function  $\delta(t-t')$ in (\ref{3})
is replaced by the unity, so that the velocity correlator
is time independent.

The RG analysis can be directly extended to this model
to prove that its Green functions also exhibit anomalous
scaling behaviour and the corresponding anomalous exponents
can be calculated in the form of series in
$\tilde\eps\equiv\eps+2$.
The critical dimensions $\Delta_{\omega}=2- \tilde\eps/2$,
$\Delta_{\theta}=-1+\tilde\eps/4$ are found exactly, and
for the structure functions defined in (\ref{HZ1}) we have
obtained:
\begin{equation}
S_{2n}\simeq D_{0}^{-n/2}\,  r^{n(2-\tilde\eps/2)}\,
(Mr)^{\tilde\Delta_{n}},
\label{400}
\end{equation}
where
\begin{equation}
\tilde\Delta_{n}=-\tilde\eps n(n-1)/(d+2)+O(\tilde\eps^{2}).
\label{401}
\end{equation}
The expressions (\ref{HZ2}) remain valid with the replacement
$ \Delta_{n}\to \tilde\Delta_{n}$.
We note that the velocity field with the dimension
$ \Delta[{\bf v}]=  1- \tilde\eps/2$ becomes dangerous for
$\tilde\eps>2$ (which corresponds to the IR divergence
of the integral in (\ref{3})
with $m=0$), and so become all its powers. However, these
operators are not Galilean invariant and therefore give no
contribution to the operator product expansions of the
structure functions and correlators (\ref{HZ2}).

\acknowledgments

The work was supported by the Russian Foundation for Fundamental
Research (grant N  96-02-17-033) and by the Grant Center for
Natural Sciences of the Russian State Committee for Higher Education
(grant N 97-0-14.1-30).

\begin{table}
\caption{Canonical dimensions of the fields and parameters in the
model (2.1).}
\label{table1}
\begin{tabular}{cccccccc}
$F$ & $\theta $ & $\theta '$ & $ {\bf v} $ & $\nu ,\nu _{0}$
& $m,M,\mu$ & $g_{0}$ & $g$ \\
\tableline
$d_{F}^{k}$ & 0 & d & -1 & -2 & 1& $\eps $ & 0 \\
$d_{F}^{\omega }$ & -1/2 & 1/2 & 1 & 1 & 0 & 0 & 0 \\
$d_{F}$ & -1 & d+1 & 1 & 0 & 1 & $\eps $ & 0 \\
\end{tabular}
\end{table}
\end{document}